\documentstyle[12pt]{article}
\title { 
{\hskip 11cm}{\normalsize{BIHEP-TH-95-38}}\\
{\bf A study on the rare radiative decay   
$B_c \rightarrow D_s^* \gamma$
}}
\author{ Dongsheng Du$^{a,b}$, Xiulin Li$^c$, Yadong Yang$^{b,d}$ 
 \thanks{E-mail address  duds@bepc3.ihep.ac.cn~~ yangyd@bepc3.ihep.ac.cn} \\
{\small $^a$ CCAST( World Laboratory ), P.O.Box 8730, Beijing, 100080, China}\\
{\small $^b$ Institute of High Energy Physics, Chinese Academy of  Sciences,}\\
{\small  P.O.Box 918(4), Beijing, 100039, China.\thanks{Mailing Address}}\\
{\small $^c$ Department of Physics, Hangzhou Teacher$^{\prime}$s 
College, Hangzhou, 310012, China}\\
{\small $^d$ Department of Physics, Henan Normal University,}\\
{\small Xinxiang, Henan, 453002, China.} }
\date{}

\begin{document}
\maketitle
\begin{abstract} 
 We study the decay $B_c \rightarrow D_s^* \gamma$. There are two
mechanisms contributing to the process. One proceeds through the short 
distance $b{\rightarrow}s\gamma$ transition and the other occurs through
weak annihilation accompanied by a photon emission.
The electromagnetic penguin contribution is estimated by perturbative
QCD and found to be $4.68\times10^{-18}GeV$.
In particular, we find the contribution  of the weak  annihilation
is $6.25\times10^{-18}GeV$ which is in the same order  as that of
the electromagnetic penguin. The total decay rate
$\Gamma(B_c \rightarrow D_s^* \gamma)$ is predicted to be
$1.45\times10^{-17}GeV$ and the branching ratio 
$Br(B_c \rightarrow D_s^* \gamma)$ is predicted to be $2.98\times 
10^{-5}$ for $\tau_{B_c}=1.35ps$.
The the decays  $B_c \rightarrow D_s^* \gamma$  
can be well studied  at LHC in the near future.\par

\end{abstract}
\newpage
\section{INTRODUCTION}
Recently the physics of $B_c$ meson have got  intensive attention[1,2].
It is believed that the $B_c$ meson is the next and the final member
of B mesons.
It provides unique  opportunities to examine various heavy quark fragmentation
models, heavy quark spin-flavor symmetry, different quarkonium bound state 
models, and properties of inclusive and exclusive decay channels.
Being made of two heavy quarks of {\it different} flavors,
$B_c$ weak decays  also offer a rich source 
to measure CKM  matrix elements of the standard
model[2].  \par
The interest of studying rare B decays lies on the fact that
those decays, induced by the flavor-changing 
neutral currents $b\rightarrow s\gamma$, 
are usually controlled  by the one-loop electromagnetic
penguin diagrams. They play important role in testing loop effects
in the standard model and in searching for  physics beyond the standard
model( so called {\it new physics}).
Most recently   the weak radiative decays of B meson and bottom baryons
have been systematically studied in [3,4].  Hence a precise study of
the weak radiative decays of $B_c$ needs to be undertaken.
In the present paper, we will address $B_c$ radiative decay
$B_c \rightarrow D_s^* \gamma $.\par

The subprocess $b\rightarrow s\gamma$, taken as a free decay, is
usually treated as  the only flavor-changing contribution to
$B\rightarrow K^* \gamma$[5], so do 
$B_c \rightarrow D_s^* \gamma$. However, boundstate effects could 
seriously modify the results of the assumption. Boundstate effects 
include modifications from weak annihilation which involves no 
flavor-changing neutral currents at all. 
The effects of the weak annihilation mechanism are expected 
to be  large in the 
$B_c$ decays.
We shall address this point in detail below.\par 
 Unfortunately, the well known chiral-symmetry[6] and the heavy quark 
 symmetry[7] cannot be applied straightforwardly to this process. 
 Recently, a perturbative QCD(PQCD) 
 analysis of the B meson exclusive decays seems give a good prediction[8].
 As it is argued in ref.[9] that $B_c$ two body nonleptonic decay 
 can be conveniently studied within the  framework of PQCD suggested by 
 Brodsky-Lepage[10] and then developed in [8].\par

 The reason to use  PQCD  to analyze of $B_c$ radiative 
 decay is as bellow. 
 In the process $b\rightarrow s\gamma$, s quark obtains   large  
 momentum by recoiling. In order to form a bound state with the spectator 
 ${\overline c}$ quark, 
 the most part of the momentum of s quark must be transferred to ${\overline c}$ by a hard   
 scattering process, since in the final bound state ({\it i.e} $D_s^{*-}$)
 the heavy charm should  share the most part of the momentum of $D_s^{*-}$. The  
 hard scattering is suitable for PQCD calculation[8,10].
 The Feynmen diagrams we  will calculate are given in Fig.1.\par
 The paper is organized as follows.
 In section 2, we display our calculations. We  present our 
 numerical results in Section 3.  Section 4  contains the discussions 
 and conclusions.\par
\section{CALCULATION}
Exclusive processes at large momentum transfer are  exploited by 
Brodsky-Lepage[10] within  PQCD starting with a Fock component
expansion of the involved hadrons, where a twist expansion
suggests that the contribution from the lowest order Fock 
component dominates
the physical observable under consideration. 
An exclusive process then involves a perturbatively 
calculable hard scattering amplitude convoluted with  nonperturbative 
soft physics wavefunctions of the 
initial and the final hadrons. Although these wavefunctions 
are incalculable from first principle, but they are universal
for each hadron, {\it i.e} they are factorized out from the  
hard scattering amplitude and hence are independent of the process involved.\par
The factorization scheme advocated by Brodsky-Lepage[10] 
is employed, 
where the momenta of quarks are taken as some fractions 
$x$ of the total momentum of the meson weighted by a soft   
physical distribution functions $\phi_H (x)$. The peaking 
approximation is used for $\phi_H (x)$[11]. The distribution  
amplitude of $B_c$ is 
$$
\phi_{B_c}(x)=\frac{1}{2\sqrt{3}}f_{B_c}\delta(x_i -\frac{m_i}{m_1 +
m_2 }),
\eqno{(1)}
$$
where $m_i$ corresponds to $m_b$, $m_c$. 
The distribution function of $D_s^{*}$ is 
$$
\phi_{D_s^*}(x)=\frac{1}{2\sqrt{3}}f_{D_s^* }\delta(x-
\frac{\epsilon_{D_s^*}}{m_{D_s^*} }),
\eqno{(2)}
$$
where $\epsilon_{D_s^*}$ is defined as
$$
 \epsilon_{D_s^*}=\frac{m_{D_s^*}-m_c}{m_{D_s^*}}.
 \eqno{(3)}
$$
 
The spinor part of $B_c$ and the final meson $D_s^*$
are
$$
 \frac{p{\hskip -2mm}/~ +m_{B_c}} {\sqrt{2}}\gamma_5,~~~~~~~~~~~~~~~~~
 \frac{p{\hskip -2mm}/~ -m_{D_s^*}} {\sqrt{2}}\epsilon{\hskip -2mm}/~,
 \eqno{(4)}
$$
where $\epsilon$ is the polarization vector of $D_s^*$.\par 

\subsection{Electromagnetic penguin contributions}  
Within the standard model, the process is governed by 
the electromagnetic penguin operators[12], for $m_s \ll m_b$
$$
H_{eff} =\frac{4G_F}{\sqrt{2}} V_{tb}V_{ts}^* C_7 (\mu) O_7 .
\eqno{(5)}
$$
Here we use the notation of Grinstein {\it et al.}[12] 
$$
O_7 =\frac{e}{16\pi^2}m_b {\overline s}\sigma^{\mu\nu}
F_{\mu\nu}\frac{1+\gamma_5}{2}b,
\eqno{(6)}
$$
we denote it by a blob in Fig.1.a.\par
The coefficient of $O_7$ includes the QCD corrections, at the scale  
of  W mass, 
$$
C_7 (m_W)=\frac{x}{24(1-x)^4}[
(1-x)(8x^2 +5x-7)+6x(3x-2)lnx],
\eqno{(7)}
$$
where $x=m_t^2 /m_W^2$, and when it  runs down to 
low scale $\mu=m_b$, it turns out to be 
$$
C_7(m_b)=y^{-\frac{16}{23}}\left[
C_{7}(m_{W})-\frac{58}{135}(y^{\frac{10}{23}}-1)
-\frac{29}{189}(y^{\frac{28}{23}}-1) \right]
\eqno{(8)}
$$
where $y=\alpha_s (m_{b})/\alpha_s (m_{W})$. We have neglected 
the mixing of $O_7$ with other operators which 
give small effects.\par
Now we write down the amplitude of Fig.1.a as
\begin{eqnarray*}
M_{a}&=&\int^1_0 dx_1 dy_1\phi^{*}_{D_s^* }\phi_{B_c } C  \\
&\times&\frac{1}{4} \left\{ 
Tr \left [
(q{\hskip -2mm}/~-m_{D_s^*})\epsilon^*{\hskip -3mm}/~
\sigma_{\mu\nu}(1+\gamma_5)k^{\nu}\eta^{*\mu}
(p{\hskip -2mm}/~-y_1 q{\hskip -2mm}/~+m_b )
\gamma_{\alpha}(p{\hskip -2mm}/~+m_{B_c})\gamma_5 
\gamma^{\alpha}
\right] \frac{1}{D_1 }\frac{1}{D_3 }  \right. \\
&+& \left.
Tr \left [
(q{\hskip -2mm}/~-m_{D_s^*})\epsilon^*{\hskip -2mm}/~
\gamma_{\alpha}
(q{\hskip -2mm}/~-x_1 p{\hskip -2mm}/~)
\sigma_{\mu\nu}(1+\gamma_5)k^{\nu}\eta^{*\mu}
(p{\hskip -2mm}/~+m_{B_c})\gamma_5 
\gamma^{\alpha}
\right] \frac{1}{D_2 }\frac{1}{D_3 } 
\right\} {\hskip 0.5cm}(9),
\end{eqnarray*}               
where $\eta$ is the polarization vector of the  photon, $x_1$ and $y_1$ are the               
momentum fractions shared by charm in $B_c$ and $D_s^*$, respectively.
The denominators are  
\begin{eqnarray*}
D_1 &=& (1-y_1 )(m_{B_c }^2 -m_{D_s^* }^2 y_1 )-m_b^2   \\
D_2 &=& (1-x_1 )(m_{D_s^* }^2  -m_{B_c }^2  x_1 )  {\hskip 5cm}(10) \\
D_3 &=&(x_1 -y_1 )(x_1 m_{B_c }^2 -m_{D_s^* }^2  y_1  ),
\end{eqnarray*}
and the constant $C$ is
$$
C=\frac{f_{B_c}}{2\sqrt{3}}
\frac{f_{D_s^*}}{2\sqrt{3}}C_F C_7 (m_b)e
\frac{\alpha_s}{4\pi}
\frac{4G_{F}}{\sqrt{2}}V_{tb}V^*_{ts} m_b.
\eqno{(11)}
$$
The result is 
$$
M_a =i\varepsilon_{\mu\nu\alpha\beta}\epsilon^{*\mu}\eta^{*\nu}
q^{\alpha}k^{\beta}f_1^{\gamma peng}  + (\eta^{*}\cdot\epsilon^{*} p\cdot k
-\eta^{*}\cdot p \epsilon^{*}\cdot k)f_2^{\gamma peng}, 
{\hskip 2cm}(12)
$$
with the formfactors
\begin{eqnarray*}               
f_1^{\gamma peng}&=&f_2^{\gamma peng}=C\int_0^1 dx_1 dy_1 \delta(x_1 
-\frac{m_c }{m_{B_c}})\delta(y_1 -1+\epsilon_{D_s^*})4 \\
&\times&
\left \{
\left [ 
m_{B_c} (1-y_1 )(m_{B_c}-2m_{D_s^*} )
-m_b (2m_{B_c}-m_{D_s^*} ) 
\right ] 
\frac{1}{D_1 }
\frac{1}{D_3 }
-m_{B_c}m_{D_s^*} (1-x_1 ) 
\frac{1}{D_2 }
\frac{1}{D_3 }
\right\},{\hskip 0.1cm}(13) 
\end{eqnarray*}               
\subsection{ The weak annihilation contribution}
Being made up of two different heavy flavors, $B_c$ meson is also 
the unique place to probe  the weak annihilation mechanism. \par
Using the formalism developed by Cheng {\it et~al. }[3], the 
amplitude of Fig.1.b  are found to be 
$$
M_b =i\varepsilon_{\mu\nu\alpha\beta}\epsilon^{*\mu}\eta^{*\nu}
q^{\alpha}k^{\beta}f_1^{anni}  + (\eta^{*}\cdot\epsilon^{*} p\cdot k
-\eta^{*}\cdot p \epsilon^{*}\cdot k)f_2^{anni}, 
{\hskip 1cm}(14)
$$
with
$$
f_1^{anni} ={\kappa}a_1 \left [ 
\left ( 
\frac{e_s }{m_s }+\frac{e_c }{m_c } 
\right )
\frac{m_{D_s^*} }{m_{B_c} }
+\left ( 
\frac{e_c }{m_c }+\frac{e_b }{m_b } 
\right )
\right ]
\frac{m_{D_s^*} m_{B_c }}{m_{B_c}^2 -m_{D_s^*}^2},
$$
$$
f_2^{anni} =-{\kappa} a_1 \left [ 
\left ( 
\frac{e_s }{m_s }-\frac{e_c }{m_c } 
\right )
\frac{m_{D_s^*} }{m_{B_c} }
+\left ( 
\frac{e_c }{m_c }-\frac{e_b }{m_b } 
\right )
\right ]
\frac{m_{D_s^*} m_{B_c }}{m_{B_c}^2 -m_{D_s^*}^2},
\eqno{(15)}
$$
where $\kappa=eG_F V_{cb}V^*_{cs}f_{B_c}f_{D^*_s}/\sqrt{2}$, and the $m_i$
is the constituent quark mass. The parameter $a_1$, corresponding to the one   
appearing in the nonleptonic $B$ decays,
includes 
the QCD corrections 
to the four  Fermion operator $({\overline c}b)_{V-A}({\overline s}c)_{V-A}$.
In nonleptonic $B$ decays, $a_1$ is extracted to be 1.01 from the recent 
CLEO data on $B\rightarrow D^{(*)}\pi(\rho)$
 and $B\rightarrow J/\Psi K^*$[13] by Cheng {\it et al.}[3]. 
 Here we take $a_1=1$ in our  
numerical results.
\section{Numerical Results} 
In order to have a numerical estimation we adopt the following parameters.\\
1. Meson mass and the constituent quark mass\\
  $$
  M_{D_s^*}=2.11GeV,~~~m_{b}=4.7GeV,~~~m_c =1.6GeV,~~~m_s =510MeV
$$                                                                 
from the Particle Data  Group[14]. $m_{B_c}$ has been  estimated to  be about 
6.27GeV in the literature[15,16]. \\ 
2. Decay constants of  $B_c$ and $D_s^*$.
Up to now  $f_{D_s}$ has been reported by  
three  groups   
\[
f_{D_s^* }=\left\{\begin{array}{cc}
232\pm45\pm20\pm48MeV,&S.Aoki~~ {\it et~ al., }[17] \\
344\pm37\pm52\pm42MeV,&D.Acosta~~ {\it et~ al., }[18] \\
430\pm^{150}_{130}\pm40MeV,~~~~ & J.Z.Bai~~{\it et~ al., }[19]
\end{array}
\right.
\]
where the first errors are statistical, the second are systematic, 
and the third  are uncertainties involved  in extracting 
the branching fraction $ B(D_s^+ \rightarrow \mu^+ \nu_{\mu})$.
 From  heavy quark symmetry,
 $f_{D_s^*}=f_{D_s}$, we take $f_{D_s^*}=344MeV$.\\
 The pseudoscalar decay constant  $f_{B_c}$ is related to 
 the ground-state $c{\overline b}$ wave function at the origin 
 by the Van Royen-Weisskopf formula[20] modified for color
 $$ 
 f_{B_c}=\sqrt{\frac{12}{m_{B_c}}}\mid \Psi_{100}(0) \mid=
\sqrt{\frac{3}{\pi m_{B_c}}}\mid R_{10}(0) \mid .
 $$
Using the recent results for $R_{10}$ derived by Quigg[21].
One finds
\[
f_{B_c }=\left\{\begin{array}{cc}
500MeV,&(Buchmuller-Tye ~~potential[22]) \\
510MeV,&(Power-law ~~potential[23]) \\ 
479MeV,&(Logarithmic ~~potential[24]) \\ 
696MeV,&(Cornell~~potential[25]) \\ 
\end{array}
\right.
\]
we will take $f_{B_c}=500MeV$ for our numerical estimation.\\
3.CKM elements and strong couplings\\
For quark mixing matrix elements, we will take $V_{cb}=0.040$[26],
$\mid V_{ts} \mid =V_{cb}$, $\mid V_{cs} \mid =0.9745 $ and $V_{tb}=1$.
The QCD coupling constant $\alpha_s(\mu)$ at any renormalization 
scale $\mu$ can be calculated from $\alpha_s(M_z )=0.117$ via
$$
\alpha_s(\mu)=\frac{\alpha_s(M_Z)}{1-\beta_0 
\frac{\alpha_s(M_Z)}{2\pi}ln(M_Z /\mu ) }
$$
and $\beta=11- \frac{2}{3}n_f$. We get $C_7 (M_W )=-0.1953 $ and
$C_7 (M_b )=-0.2939 $. \par
With the values given above for various quantities, we 
can compute the form factors, the decay rates and the branching 
ratios.\par
The results are
$$
f_1^{\gamma peng}=1.170\times 10^{-9}, ~f_2^{\gamma peng} =f_1^{\gamma peng},~~
f_1^{anni} =1.724\times 10^{-9}, ~~~f_2^{anni} =-8.273\times 10^{-10}, 
$$
in units of $GeV^{-1}$.\\
The ratios of long- and short distance contributions to the form factors  
$f_1$ and $f_2$ are 
$$
\frac{f_1^{anni}}{f_1^{\gamma peng}}=1.474,~~~~~
\frac{f_2^{anni}}{f_2^{\gamma peng}}=-0.710,
$$
which are independent of $f_{B_c}$ and $f_{D_s^*}$, and expected to be 
reliable. \par
From the amplitude formula eq(12)(14), we get   
$$
\Gamma(B_{c}\rightarrow D_s^* \gamma)=
\frac{ (m_{B_c}^2 -m_{D_s^* }^2 )^3 }{32\pi m_{B_c}^3}
[ f_1^2 +f_2^2 ].
$$
We finally obtain the decay rates 
\[
\Gamma(B_c \rightarrow D_s^* \gamma)=
 \left\{\begin{array}{cc}
 4.68\times10^{-18}GeV, & only ~~penguin\\ 
 6.25\times10^{-18}GeV, & only ~~annihilation\\
 1.45\times10^{-17}GeV, & penguin + annihilation
 \end{array}
\right.
\] 
The lifetime of $B_c$ is predicted to be $\tau_{B_c}=1.35\pm .15ps$ 
by Quigg[27], but other theorists report smaller values   
$\tau_{B_c}=0.4ps\sim 0.7ps$[27]. We estimate the branching ratio     
 $Br(B_c \rightarrow D_s^* \gamma)$ 
as a function of $\tau_{B_c}$. The results are given in Table 1. 
\begin{center}
\begin{tabular}{|c|c|c|c|c|}\hline
$Br(B_{c}{\rightarrow}D_s^* \gamma)$ & 
 $\tau_{B_c}=0.4ps$ & $\tau_{B_c}=0.7ps$ &  $\tau_{B_c}=1.0ps$ & 
  $\tau_{B_c}=1.35ps$\\ \hline
$Br^{peng}$ & 2.85$\times10^{-6}$ &4.98$\times10^{-6}$ &7.11$\times10^{-6}$
& 9.61$\times10^{-6}$ \\  \hline
$Br^{anni}$ & 3.80$\times10^{-6}$ & 6.65$\times10^{-6}$ & 
9.50$\times10^{-6}$& 1.28$\times10^{-5}$ \\  \hline
$Br^{total}$  & 8.83$\times10^{-6}$ & 1.54$\times10^{-5}$ & 
2.21$\times10^{-5}$&2.98$\times10^{-5}$ \\  \hline
\end{tabular} \\
Table.1
\end{center}
 \section{DISCUSSION AND CONCLUSION}
We have studied two mechanism contributing to the process 
$B_c \rightarrow D_s^* \gamma$. The short distance one(Fig.1.a) 
induced by electro-magnetic penguin has been estimated within PQCD 
framework which is suitable for the process involved large 
energy release and hard scattering. The distribution functions 
$\phi_H (x)$  coincide with the usual nonrelativistic Coulomb
wave functions of $c{\overline b}$ and $c{\overline s}$ systems, 
and  expected to work well in  the present case.\par
An unique nature of $B_c$ decays is that "the spectator"  is heavy.
Assume that the velocities of the heavy spectator 
are equal to $v_1$ of $B_c$(before scattering ) and to $v_2$  of the 
final meson(after scattering ).  In  two body $B_c$ decays,     
$v_1 \cdot v_2$ is always  large. For $B_c \rightarrow D_s^* \gamma$, 
$v_1 \cdot v_2 =1.7$. Certainly, there is no phase-space for the 
propagators appearing in Fig.1.a to be on-shell, so 
the imaginary   
part of $M_a$ is absent. This is different from the situations in ref.[8].
Here we recall that the momentum square of the hard scattering
exchanged by gluon is about $3.6GeV^2$, which is large enough  
for PQCD analysis. The hard scattering process cannot be 
included conveniently in the soft hadronic process described 
by the wavefunction of the final bound state, that is an important
reason why we cannot apply 
 the commonly used models  with {\it spectator ansatz}, 
for example, HQET[7], BSW, ISGW[28] models,    
to the two body  $B_c$ decays. That is also one of the reasons why 
the commonly used models  cannot reliably predict the processes of  
$B\rightarrow$ two light hadrons or $\gamma$+light hadron, since 
the large mass scale  of the decaying $B$ meson and the nearly massless 
final states imply that the transition amplitude is governed by hard process.  
Hence we conclude that the {\it spectator ansatz} is broken down in 
such kind of process and it is misleading to apply straightforwardly the 
 commonly used models  with {\it spectator ansatz} to such 
 processes  discussed above.
Another competing mechanism is the weak annihilation. We find that 
it is equally  important as the former ones. 
The situation  is different from that of the radiative weak $B^{\pm}$ decays
which are overwhelingly dominated by  electromagnetic penguin contributions.
The results stem from at least two reasons:\\
1. The compact size of $B_c$ meson enhances the importance of the 
annihilation
decays.\\
2. In $B_c \rightarrow D_s^* \gamma $ case,
the CKM  factor of weak annihilation contribution is  
$\mid V_{cb}V_{cs}\mid$, but in  $B^{\pm} \rightarrow K^* \gamma $ case,
the CKM part is $\mid V_{ub}V_{us}\mid$ which is much smaller
than the former.\par
We  have neglected the contribution of the vector meson dominance(VMD).
Using the methods developed in [29], combined with the PQCD calculation  
of the  form factors of the process $B_c \rightarrow D_{s}^{*} 
J/\Psi(\Psi^{\prime}) $ (following the conversion $J/\Psi(\Psi^{\prime})
\rightarrow \gamma$ ), we find that the VMD contribution is negligibly  
small,due to the small  $J/\Psi(\Psi^{\prime})\rightarrow \gamma$
coupling   $e/g_{J/\Psi(\Psi^{\prime}) \gamma}\approx$0.025(0.016). \par
Finally, we want to say a few words about the possibility of observing the interesting process 
at Tevatron and at the CERN Large Hadron Collider(LHC).
The numbers of $B_c$ produced at Tevatron and LHC are estimated 
to be [30]                                          
16000 (for $25Pb^{-1}$ integrated luminosities 
with cuts of $P_T (B_c ) > 10GeV, {\mid}y(B_c ){\mid}<1$ )
and
2.1$\times 10^8$ (for $100fb^{-1}$ integrated luminosities 
with cuts of $P_T (B_c ) > 20GeV, {\mid}y(B_c ){\mid}< 2.5$ ),
respectively. Using the numbers and $Br(B_c  \rightarrow D_s^* \gamma)
$, we find that the decay channel is unobservable 
at Tevatron, but more than one thousand events of interest will 
be produced  at LHC, so it can be well studied  at LHC in the near future.\par
\section{Acknowledgment}
This work is supported in part by the National Natural Science Foundation 
and the Grand of State Commission of Science and Technology of China.
\newpage
\begin{center}
{\bf Reference}
\end{center}
\begin{enumerate}

\item 
Dongsheng Du and Zi Wang, Phys.Rev.D39(1989)1342; \\
K.Cheung, T.C.Yuan, Phys.Lett.B325(1994)481.
E.Braaten, K.Cheung,T.C.Yuan,\\ 
Phys.Rev.D48(1994)R5049,{\it ibid} 48(1993)4230. 
\item
L.G.Lu, Y.D.Yang, and H.B.Li, Phys.Lett.B341(1995);Phys.Rev.D51(1995)2201; 
\item
H.Y.Cheng, C.Y.Cheung, G.L.Lin, {\it et~al.,} Phys.Rev.D51(1995)1199.
\item
A.J.Buras, M.Misiak, {\it et~al.,} Nucl.Phys.B424(1994)374; B.Grinstein, M.Springer
and M.B.Wise, Nucl.Phys.B339(1990)269l; M.Misiak, Phys.Lett.B269(1991)169.
\item
J.Tang, J.H.Liu and K.T.Chao, Phys.Rev.D51(1995)3051; \\
K.C.Bowler {\it et~al.,} Phys.Rev.Lett72(1994)1398.
\item
H.Leutwyler and M.Roos, Z.Phys.C25(1984)91.
\item
M.Neubert, Phys.Rep.245(1994)1398; and reference therein.
\item
A.Szczepaiak, E.M.Henley, and S.J.Brodsky, Phys.Lett.B243(1990)287.
C.E.Carlson and J.Milana, Phys.Lett.B301(1993)237, Phys.Rev.D49(1994)5908, 
ibid 51(95)450;
H-n.Li and H.L.Yu, Phys.Rev.Lett74(1995)4388.
\item
Dongsheng Du, Gongru Lu, and Yadong Yang, BIHEP-TH-32(submitted to  
Phys.Lett.B).
\item
S.J.Brodsky and G.P.Lepage, Phys.Rev.D22(1980)2157; for a more recent review 
see{\it Perturbative QCD}, edited by A.H.Mueller (World Scientific, 
Singapore, 1989).
\item
S.J.Brodsky and C.R.Ji, Phys.Rev.Lett55(1985)2257.
\item
B.Grinstein, R.Springer,  and M.B.Wise, Nucl.Phys.B339(1990)269;
\item
CLEO Collaboration, M.S.Alam {\it et~al.,} Phys.Rev.D50(1994)43.
\item
Particle Data Group, L.Montanet  {\it et~al.,} Phys.Rev.D50(1994)1173.
The constituent light quark masses are given on 1729.
\item
W.Kwong and J.L.Rosner, Phys.Rev.D44(1991)212.
\item
E.J.Eichten and C.Quigg FERMILAB-PUB-94/032-T, hep-ph/942210.
\item
A.Aoki {\it et~al.,} Prog.Theor.Phys 89(1993)137.
\item
D.Acosta {\it et~al.,} CLNS93/1238 ,
\item
J.Z.Bai {\it et~al.,} BES Collaboration Phys.Rev.Lett74(1995)4599.
\item
R.Van Royen and V.F.Weisskopf, Nuovo Cim50(1967); 51(1967)583.
\item
C.Quigg Phys.Rev.D52(1995)1726.
\item
W.Buchm\"uller and S-H.H.Tye, Phys.Rev.D24(1994)24(1981)132.
\item
A.Martin, Phys.Lett.B93(1980)338.
\item
C.Quigg and J.L.Rosner, Phys.Lett.B71(1977)153.
\item
E.Eichten, K.Gottfried, T.Kinoshita, K.D.Lane and T.M.Yan, Phys.Rev.D17(1978)3090;
ibid 21(1980)203.
\item
M.Neubert, Phys.Lett.B238(1994)84.
\item
C.Quigg, FERMILAB-Conf-93/265-T;
C.H.Chang and Y.Q.Chen, Phys.Rev.D49(1994)3399;
P.Colangelo, {\it et~al.,} Z.Phys.C57(1993)43;\\
S.S.Greshtein,{\it et~al.,} Int.J.Mod.Phys.A6(1991)2309.
\item
M.Bauer, B.Stech and M.Wirble, Z.Phys.C29(1985)637, ibid C34(1987)1;
B.Grinstein, N.Isgur, D.Scora and Wise Phys.Rev.D39(1987)799;
\item
E.Golowich and S.Pakvasa Phys.Rev.D51(1995)1215;
H.Y.Cheng Phys.Rev.D51(1995)6228.
\item
K.Cheung Phys.Rev.Lett71(1993)3413.

\end{enumerate}
 
\newpage
\begin{center}
{\bf Table and Figure Captions }
\end{center}
Table.1. $Br(B_{c}{\rightarrow}D_s^* \gamma)$ is displayed for varous 
$B_c$ lifetime. 
The second raw is for the contribution of electromagnetic penguin. 
The third  raw is for the contribution of annihilation.
The final line is for the total results.
Fig.1.a. The leading penguin graphs contribute to 
$B_c{\rightarrow}D_s^* \gamma$.\\ 

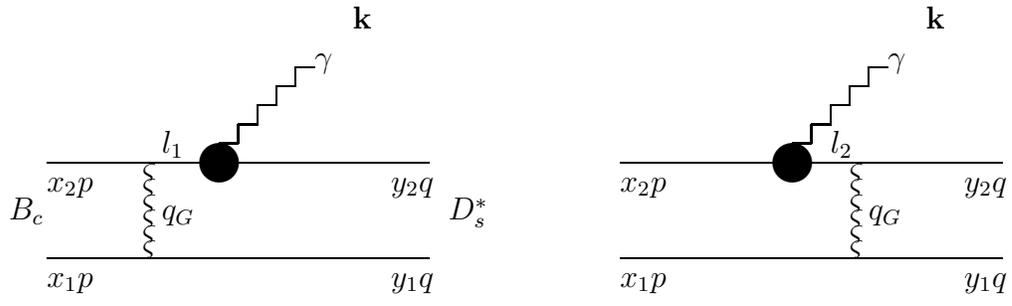
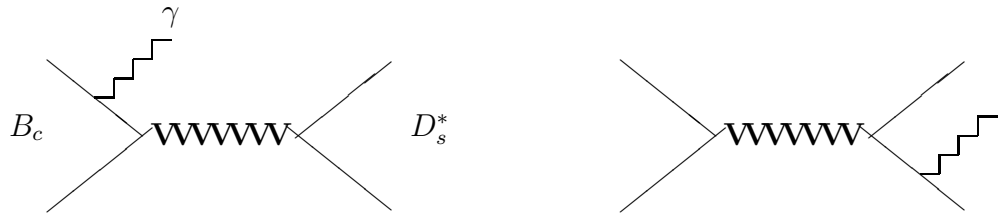
Fig.1.b. The leading W-annihilations to $B_c{\rightarrow}D_s^* \gamma$. 
Contributions due  to photon emission from other quarks are denoted by 
ellipses.

\newpage
 \begin{picture}(30,0)
{\bf 
\setlength{\unitlength}{0.1in}
\put(5,-15){\line(1,0){20}}
\put(5,-20){\line(1,0){20}}
 \put(3,-18){$B_c$}  
  \put(26,-18){$D_s^*$}
   \multiput(10,-15.8)(0,-0.8){6}{$\varsigma$}
   \multiput(14,-15)(1,1){5}{\line(0,1){1}} 
   \multiput(13,-15)(1,1){6}{\line(1,0){1}}
    \put(14,-15){\circle*{3}}
      \put(11,-18){$q_G$}
\put(5,-16.5){$x_2 p$}
\put(5,-21.5){$x_1 p$}
\put(23,-16.5){$y_2 q$}
\put(23,-21.5){$y_1 q$}
\put(11,-14.5){$l_1 $}
\put(29,-25.2){(a)}
\put(21,-8){k}
\put(19,-10){$\gamma$} 
\put(35,-15){\line(1,0){20}}
\put(35,-20){\line(1,0){20}}
   \multiput(47,-15.8)(0,-0.8){6}{$\varsigma$}
   \multiput(44,-15)(1,1){5}{\line(0,1){1}} 
   \multiput(43,-15)(1,1){6}{\line(1,0){1}}
\put(48,-18){$q_G$}
\put(35,-16.5){$x_2 p$}
\put(35,-21.5){$x_1 p$}
\put(53,-16.5){$y_2 q$}
\put(53,-21.5){$y_1 q$}
\put(46,-14.5){$l_2 $}
\put(44,-15){\circle*{3}} 
\put(51,-8){k}
\put(49,-10){$\gamma$}
\put(5,-35){\line(5,-4){5}}
\put(5,-43){\line(5,4){5.5}}
\multiput(10.4,-39.5)(1,0){7}{V}
\put(17.5,-38.5){\line(5,-4){5.5}}
\put(18,-39.1){\line(5,4){5}}
   \multiput(8.5,-37)(1,1){3}{\line(0,1){1}} 
   \multiput(7.5,-37)(1,1){4}{\line(1,0){1}}
 \put(3,-39){$B_c$}  
  \put(24,-39){$D_s^*$}
\put(11,-33){$\gamma$}
\put(35,-35){\line(5,-4){5}}
\put(35,-43){\line(5,4){5.5}}
\multiput(40.4,-39.5)(1,0){7}{V}
\put(47.5,-38.5){\line(5,-4){5.5}}
\put(48,-39.1){\line(5,4){5}}
\multiput(51.7,-41)(1,1){3}{\line(0,1){1}} 
\multiput(50.7,-41)(1,1){4}{\line(1,0){1}}
\multiput(25,-52)(1.5,0){6}{$\bullet$}
\put(29,-60){(b)} 
\put(29,-78){Fig.1} 
}
\

\end{picture}
\end{document}